\documentclass[12pt]{article}
\usepackage{amsmath}
\usepackage{graphicx}
\usepackage{subfigure}
\usepackage{hyperref}

\begin{document}

\author{Ernst Trojan \\ \textit{Moscow Institute of Physics and Technology}}
\title{Specific heat and entropy of tachyon Fermi gas}
\maketitle

\begin{abstract}
We consider an ideal Fermi gas of tachyons and derive a low temperature
expansion of its thermodynamical functions. The tachyonic specific heat is
linear dependent on temperature $C_V=\varepsilon _Fk_FT$ and formally
coincides with the specific heat of electron gas if the tachyon Fermi energy
is defined as $\varepsilon _F=\sqrt{k_F-m^2}$.
\end{abstract}


\section{Introduction}

The concept of tachyon fields plays significant role in the modern research,
and tachyons are considered as candidates for the dark matter and dark
energy, they often appear in the brane theories and cosmological models.
Tachyons, are commonly known as instabilities with energy spectrum 
\begin{equation}
\varepsilon _k=\sqrt{k^2-m^2}\qquad k>m  \label{t}
\end{equation}
where $m$ is the tachyon mass and relativistic units $c=\hbar =1$\ are used.

A system of many tachyons can be studied in the frames of statistical
mechanics \cite{M84,DHR89}, and thermodynamical functions of ideal tachyon
Fermi and Bose gases are calculated \cite{KRS07}. We have recently studied
the equation of state (EOS) and acoustic properties of the cold tachyon
Fermi \cite{TV2011c} and Fermi gas of tachyonic thermal excitations \cite
{TV2011d}.

In the present paper we consider a Fermi gas of tachyons at finite but low
temperature. When the temperature $T$ is much lower than the Fermi energy $%
\varepsilon _F$ of cold tachyon gas at zero temperature, all thermodynamical
functions are expanded into a series of $T/\varepsilon _F$. We want to find
such important quantities as the entropy and specific heat (heat capacity)
of tachyon Fermi gas. This problem has been already initiated, however, for
the limiting nonrelativistic and ultrarelativistic cases \cite{KRS07b}, and
we proceed with its general solution for arbitrary range of parameters.

\section{ Tachyon Fermi gas}

Consider a system of free tachyons with the energy spectrum $\varepsilon _k$
(\ref{t}) that obey the Fermi statistics. Its energy density $E$ and
particle number density $n$ are defined by standard formulas~\cite
{Kapusta89,TV2011c}: 
\begin{equation}
E=\frac \gamma {2\pi ^2}\int\limits_m^\infty \varepsilon _k\,f_k\,k^2dk
\label{e}
\end{equation}
\begin{equation}
n=\frac \gamma {2\pi ^2}\int\limits_m^\infty f_k\,k^2dk  \label{n}
\end{equation}
where $\gamma $ is the degeneracy factor, and 
\begin{equation}
f_k=\frac 1{\exp \left[ \left( \varepsilon _k-\mu \right) /T\right] +1}
\label{f}
\end{equation}
is the distribution function, while $\mu $ is the chemical potential of
tachyon Fermi gas at temperature $T$. The specific heat is also determined
by standard formula \cite{LL5} 
\begin{equation}
C_V=T\frac{\partial S}{\partial T}=\frac{\partial E}{\partial T}  \label{c}
\end{equation}
where $S$ is the entropy density.

If we introduce dimensionless variables 
\begin{equation}
x=\frac{\varepsilon _k\,}T\qquad \beta =\frac mT\qquad \lambda =\frac \mu T
\label{dim}
\end{equation}
the thermodynamical functions of tachyons (\ref{e})-(\ref{n}) will be
written so 
\begin{equation}
E=\frac{\gamma T^4}{2\pi ^2}\int\limits_0^\infty \frac{\sqrt{x^2+\beta ^2}%
x^2dx}{\exp \left( x-\lambda \right) +1}  \label{e1}
\end{equation}
\begin{equation}
n=\frac{\gamma T^3}{2\pi ^2}\int\limits_0^\infty \frac{\sqrt{x^2+\beta ^2}xdx%
}{\exp \left( x-\lambda \right) +1}  \label{n1}
\end{equation}
Both integrals can be also presented in the following universal form 
\begin{equation}
Q=\sigma \left( T\right) J\left( \lambda \right)   \label{ie}
\end{equation}
where $\sigma \left( T\right) $ is a function of temperature, while integral

\begin{equation}
J\left( \lambda \right) =\int\limits_0^\infty g\left( x\right) f_k\left(
x,\lambda \right) dx  \label{i}
\end{equation}
includes the distribution function $f_k$ (\ref{f}) and function 
\begin{equation}
g\left( x\right)  \label{g}
\end{equation}
each taken for the energy density and particle number density.

At zero temperature the chemical potential $\mu \rightarrow \varepsilon _F$
tends to the Fermi energy 
\begin{equation}
\varepsilon _F=\sqrt{k_F^2-m^2}  \label{fer}
\end{equation}
corresponding to the Fermi momentum $k_F$. The distribution function of
fermions (\ref{f}) degenerates into the Heaviside step-function 
\begin{equation}
f_k=\Theta \left( \varepsilon _F-\varepsilon _k\right)   \label{hv}
\end{equation}
Then, the energy and particle number density of tachyon Fermi gas are
immediately calculated \cite{TV2011c} 
\begin{equation}
E_0=\frac \gamma {2\pi ^2}\int\limits_m^{k_F}\sqrt{k^2-m^2}k^2dk=\frac
\gamma {8\pi ^2}\left[ k_F^3\varepsilon _F-\frac 12m^2\left( k_F\varepsilon
_F+m^2\ln \frac{k_F+\varepsilon _F}m\right) \right]   \label{e0}
\end{equation}
\begin{equation}
n=\frac \gamma {2\pi ^2}\int\limits_m^{k_F}\sqrt{k^2-m^2}kdk=\frac \gamma
{6\pi ^2}\left( k_F^3-m^3\right)   \label{n0}
\end{equation}
The latter formula determines the Fermi momentum of tachyons at zero
temperature 
\begin{equation}
k_F=\left( \frac{6\pi ^2n}\gamma +m^3\right) ^{1/3}  \label{k0}
\end{equation}
Now we are looking for low-temperature corrections to formulas (\ref{e0})-(%
\ref{n0}).

\section{Low temperature expansion}

At low temperature 
\begin{equation}
T\ll \varepsilon _F  \label{t2}
\end{equation}
the chemical potential $\mu $ is close to the Fermi energy $\varepsilon _F$ (%
\ref{fer}). However, the thermodynamical functions depend on the temperature
so that the Fermi gas has finite entropy. In order to calculate the specific
heat of tachyon Fermi gas (\ref{c}) we need a low-temperature expansion of
thermodynamical functions \cite{Ziman,TV2011f}.

Integrating (\ref{i}) by parts, we have 
\begin{equation}
J\left( \lambda \right) =\left. G\left( x\right) f_k\left( x\right) \right|
_0^\infty -\int\limits_0^\infty G\left( x\right) f_k^{\prime }\left(
x\right) dx  \label{i1}
\end{equation}
where 
\begin{equation}
f_k^{\prime }\left( x\right) =\frac{\partial f_k\left( x\right) }{\partial x}%
=-\frac{\exp \left( x-\lambda \right) }{\left[ \exp \left( x-\lambda \right)
+1\right] ^2}  \label{fpr}
\end{equation}
and 
\begin{equation}
G\left( x\right) =\int g\left( x\right) dx  \label{G}
\end{equation}

According to (\ref{f}), the distribution function has the following
asymptotic behavior 
\begin{equation}
f_k\left( 0\right) =1\qquad \underset{x\rightarrow \infty }{\lim }%
f_k\left( x\right) \sim \,\underset{x\rightarrow \infty }{\lim }\exp
\left( \lambda -x\right) =0  \label{f0}
\end{equation}
Hence 
\begin{equation}
J_0=\left. G\left( x\right) f_k\left( x\right) \right| _0^\infty =%
\underset{x\rightarrow \infty }{\lim }\left[ G\left( x\right) f_k\left(
x\right) \right] -\underset{x\rightarrow 0}{\lim }\left[ G\left( x\right)
f_k\left( x\right) \right] =-G\left( 0\right)  \label{i0}
\end{equation}
Therefore, integral (\ref{i1}) is immediately written in the form 
\begin{equation}
J\left( \lambda \right) =-G\left( 0\right) -\int\limits_0^\infty G\left(
x\right) f_p^{\prime }\left( x\right) dx  \label{i11}
\end{equation}

Let us expand function $G\left( x\right) $ in a Taylor series 
\begin{equation}
G\left( x\right) =G\left( \lambda \right) +\sum\limits_{k=1}^{k=\infty }%
\frac{g^{\left( k-1\right) }\left( \lambda \right) }{k!}\left( x-\lambda
\right) ^k  \label{tay}
\end{equation}
where 
\begin{equation}
g^{\left( k\right) }\left( x\right) =\frac{\partial ^kg\left( x\right) }{%
\partial x^k}  \label{gk}
\end{equation}

Substituting (\ref{tay}) in (\ref{i11}) we have 
\begin{equation}
J\left( \lambda \right) =-G\left( 0\right) -G\left( \lambda \right)
\int\limits_0^\infty f_k^{\prime }\left( x\right)
dx-\sum\limits_{k=1}^{k=\infty }\frac{g^{\left( k-1\right) }\left( \lambda
\right) }{k!}\int\limits_0^\infty \left( x-\lambda \right) ^kf_p^{\prime
}\left( x\right) dx  \label{i2}
\end{equation}
In the light of (\ref{f0}), the first term in (\ref{i2}) is simplified so 
\begin{equation}
-G\left( \lambda \right) \int\limits_0^\infty f_k^{\prime }\left( x\right)
dx=-\left. G\left( \lambda \right) f_k\left( x\right) \right| _0^\infty
=G\left( \lambda \right)  \label{i22}
\end{equation}
Hence 
\begin{equation}
J\left( \lambda \right) =-G\left( 0\right) +G\left( \lambda \right)
-\sum\limits_{k=1}^{k=\infty }\frac{g^{\left( k-1\right) }\left( \lambda
\right) }{k!}\int\limits_0^\infty \left( x-\lambda \right) ^kf_p^{\prime
}\left( x\right) dx  \label{i3}
\end{equation}
where $f_p^{\prime }\left( x\right) $ is determined by (\ref{fpr}). At low
temperature ($\lambda \gg 1$) integral (\ref{i3}) is approximated by formula 
\begin{equation}
J\left( \lambda \right) \cong G\left( \lambda \right) -G\left( 0\right)
+\sum\limits_{k=1}^{k=\infty }g^{\left( k\right) }\left( \lambda \right) C_k
\label{i4}
\end{equation}
with coefficients 
\begin{eqnarray}
C_k &=&\frac 1{k!}\int\limits_0^\infty \left( x-\lambda \right) ^{2k}\frac{%
\exp \left[ \left( x-\lambda \right) \right] }{\left\{ \exp \left[ \left(
x-\lambda \right) \right] +1\right\} ^2}dx=  \nonumber \\
&& \quad =\int\limits_{-\lambda }^\infty x^{2k}\frac{\exp \left( x\right) }{%
\left( \exp \left( x\right) +1\right) ^2}dx\cong \int\limits_{-\infty
}^\infty x^k\frac{\exp \left( x\right) }{\left( \exp \left( x\right)
+1\right) ^2}dx  \label{co}
\end{eqnarray}
where all odd coefficients (\ref{co}) are tending to zero 
\begin{equation}
C_{2k+1}\rightarrow 0  \label{co1}
\end{equation}
Integral (\ref{i4}) can be written in the explicit form 
\begin{equation}
J\left( \lambda \right) =G\left( \lambda \right) -G\left( 0\right)
+g^{\prime }\left( \lambda \right) \frac{\pi ^2}6+g^{\prime \prime \prime
}\left( \lambda \right) \frac{7\pi ^4}{360}+...  \label{i55}
\end{equation}
Formula (\ref{i55}) determines a low temperature expansion of
thermodynamical quantity (\ref{ie}) corresponding to function $g$ (\ref{g}).

\section{Thermodynamical functions}

Let us find a low-temperature expansion of the particle number density (\ref
{n1}), whose presentation according to (\ref{i}) includes 
\begin{equation}
\sigma _n\left( T\right) =\frac{\gamma T^3}{2\pi ^2}  \label{is}
\end{equation}
\begin{equation}
g_n\left( x\right) =x^{1/2}  \label{g1}
\end{equation}
so that the relevant function (\ref{G}) will be \textrm{\ } 
\begin{equation}
G_n\left( x\right) =\frac 13\sqrt{\left( x^2+\beta ^2\right) ^3}  \label{g2}
\end{equation}
Substituting function (\ref{g1}) and (\ref{g2}) in integral (\ref{i55}), we
obtain, up to the second-order terms: 
\begin{equation}
J_n\left( \lambda \right) =\frac{1}3 \sqrt{\left( \lambda ^2+\beta ^2\right)
^3}-\frac{1}3{\beta ^3}+\frac{\pi ^2}6\frac{2\lambda ^2+\beta ^2}{\sqrt{%
\lambda ^2+\beta ^2}}  \label{i6}
\end{equation}

Substituting (\ref{is}) and (\ref{i6}) in (\ref{i}) we find the particle
number density 
\begin{equation}
n=\frac \gamma {6\pi ^2}\left( q^3-m^3\right) +\frac \gamma {12}\frac{%
2q^2-m^2}qT^2  \label{nz}
\end{equation}
where 
\begin{equation}
q=\sqrt{\mu ^2+m^2}>m  \label{q}
\end{equation}
is the Fermi momentum at low temperature.

At zero temperature limits $\mu \rightarrow \varepsilon _F$ and $%
q\rightarrow k_F$ takes place, and formula (\ref{nz}) is reduced to (\ref{n0}%
). Hence, the Fermi momentum at low temperature is approximated by formula

\begin{equation}
q=k_F\left( 1-\frac{\pi ^2}6\frac{k_F^2+\varepsilon _F^2}{k_F^4}T^2\right)
\label{qz}
\end{equation}
and the Fermi level is 
\begin{equation}
\mu =\varepsilon _F\left( 1-\frac{\pi ^2}6\frac{k_F^2+\varepsilon _F^2}{%
k_F^2\varepsilon _F^2}T^2\right)  \label{mz}
\end{equation}

Now let us find a low-temperature expansion of the energy density (\ref{e1}%
), which is presented in the form of (\ref{i}) with 
\begin{equation}
\sigma _E\left( T\right) =\frac{\gamma T^4}{2\pi ^2}  \label{ise}
\end{equation}
\begin{equation}
g_E\left( x\right) =x^2\sqrt{x^2+\beta ^2}  \label{g1e}
\end{equation}
so that the relevant function (\ref{G}) will be \textrm{\ } 
\begin{equation}
G_E\left( x\right) =\frac 14x^3\sqrt{x^2+\beta ^2}+\frac 18\beta ^2x\sqrt{%
x^2+\beta ^2}-\frac 18\beta ^4\ln \left( x+\sqrt{x^2+\beta ^2}\right)
\label{g2e}
\end{equation}
Substituting function (\ref{g1e})-(\ref{g2e}) in integral (\ref{i55}), we
obtain

\begin{equation}
J_E\left( \lambda \right) =\frac 14\lambda ^3\sqrt{\lambda ^2+\beta ^2}%
+\frac 18\beta ^2\lambda \sqrt{\lambda ^2+\beta ^2}-\frac 18\beta ^4\ln 
\frac{\lambda +\sqrt{\lambda ^2+\beta ^2}}\beta +\frac{\pi ^2}6\lambda \frac{%
3\lambda ^2+2\beta ^2}{\sqrt{\lambda ^2+\beta ^2}}  \label{i7}
\end{equation}
Substituting (\ref{ise}) and (\ref{i7}) in (\ref{i}) we define the energy
density 
\begin{equation}
E=\frac \gamma {8\pi ^2}q^3\mu -\frac \gamma {16\pi ^2}m^2\left( \mu
q+m^2\ln \frac{\mu +q}m\right) +\frac \gamma {12}\mu \frac{\mu ^2+2q^2}qT^2
\label{ez0}
\end{equation}
Substituting (\ref{qz}) and (\ref{mz}) in (\ref{ez0}) we find

\begin{equation}
E=E_0+\frac \gamma {12}\varepsilon _Fk_FT^2  \label{ez}
\end{equation}
where $E_0$ (\ref{e0}) is the energy density of tachyon Fermi gas at zero
temperature.

Thus, according to (\ref{c}) and (\ref{ez}), the tachyonic specific heat is

\begin{equation}
C_V=\frac \gamma 6\varepsilon _Fk_FT  \label{c1}
\end{equation}
and, according to formulas (\ref{c}) and (\ref{c1}) the entropy density of
tachyons is 
\begin{equation}
S=\frac \gamma 6\varepsilon _Fk_FT  \label{s}
\end{equation}

\section{Conclusion}

The energy density of tachyon Fermi gas at low temperature $T\ll \varepsilon
_F$ is determined by formula (\ref{ez}). The specific heat of tachyons (\ref
{c1}) and the tachyonic entropy (\ref{c1}) are linear dependent on
temperature that bear resemblance with ordinary Fermi gas of subluminal
particles.

Moreover, expression for tachyonic specific heat (\ref{c1}) formally
satisfies the general formula for nonrelativistic electronic specific heat 
\cite{Ziman2}\textrm{\ } 
\begin{equation}
C_V=\frac{\pi ^2}3N\left( \varepsilon _F\right) T  \label{c2}
\end{equation}
where the density of states 
\begin{equation}
N\left( \varepsilon _k\right) =\frac \gamma {2\pi ^2}\sqrt{\varepsilon
_k^2+m^2}\varepsilon _k  \label{N}
\end{equation}
is incorporated in formula\textrm{\ \ } 
\begin{equation}
n=\int\limits_0^\infty \frac{N\left( \varepsilon _k\right) d\varepsilon _k}{%
\exp \left[ \left( \varepsilon _k-\mu \right) /T\right] +1}  \label{N2}
\end{equation}

The Fermi level of tachyon gas at finite temperature is shifted with respect
to the zero-temperature level (\ref{mz}), This formula for ultrarelativistic
tachyons at large particle number density $n$ ($k_F\gg m$ ) implies 
\begin{equation}
\mu \cong \varepsilon _F\left( 1-\frac{\pi ^2}3\frac{T^2}{\varepsilon _F^2}%
\right)  \label{mz2}
\end{equation}
while for non-relativistic tachyons at small density $n$ ($k_F\rightarrow m$
) it will be 
\begin{equation}
\mu \cong \varepsilon _F\left( 1-\frac{\pi ^2}6\frac{T^2}{\varepsilon _F^2}%
\right)  \label{mz3}
\end{equation}
where the Fermi energy of non-relativistic tachyons, according to (\ref{fer}%
) and (\ref{k0}), is defined so 
\begin{equation}
\varepsilon _F=\sqrt{\frac{4\pi ^2n}\gamma }\ll m  \label{fern}
\end{equation}
Formulas (\ref{mz2}) and (\ref{mz3}) coincide with the relevant expressions
obtained in the earlier research \cite{KRS07b}. Formula (\ref{mz}) is the
general expression for arbitrary relation between $k_F$ and $m$, while
formulas (\ref{mz3})-(\ref{fern}) are applied only at low density when $%
k_F\rightarrow m$.

However, we should not forget that tachyon Fermi gas at low density is
unstable to the causality \cite{TV2011c}, and its Fermi momentum must exceed
critical value 
\begin{equation}
k_F>k_T=\sqrt{\frac 32}m  \label{ca}
\end{equation}
that corresponds to 
\begin{equation}
\varepsilon _F>\varepsilon _T=\frac m{\sqrt{2}}  \label{ca2}
\end{equation}
It is the minimum possible Fermi energy of stable tachyon gas at zero
temperature. At finite temperature, according to (\ref{mz}), it will be 
\begin{equation}
\mu _T=\varepsilon _T\left( 1-\frac{2\pi ^2}9\frac{T^2}{\varepsilon _T^2}%
\right)  \label{mt}
\end{equation}
that also much more sufficient than the non-relativistic energy shift (\ref
{mz3}).

Since the Fermi level of stable tachyon gas must be always higher than $%
\varepsilon _T$ (\ref{ca2}), we can warrant that temperature\textrm{\ } 
\begin{equation}
T<\frac{\varepsilon _T}3\sim \frac m4  \label{ca3}
\end{equation}
should be regarded as definitely small at all $k_F>k_T$\textrm{.}

The author is grateful to Erwin Schmidt for helpful conversations.


\begin{thebibliography}{99}
\bibitem{M84}  St. Mr\'owczy\'nski, Nuovo Cim. B \textbf{81}, 179 (1984).

\bibitem{DHR89}  R. L. Dawe, K. C. Hines and S. J. Robinson, Nuovo Cim. A 
\textbf{101}, 163 (1989).

\bibitem{KRS07}  K. Kowalski, J. Rembielinski, and K.A. Smolinski, Phys.Rev.
D \textbf{76}, 045018 (2007). {\tt arXiv:0712.2725 [hep-th]}

\bibitem{TV2011c}  E. Trojan and G. V. Vlasov, Phys. Rev. D \textbf{83},
124013 (2011). {\tt arXiv:1103.2276 [hep-ph]}

\bibitem{TV2011d}  E. Trojan and G. V. Vlasov, Tachyonic thermal excitations
and causality. {\tt arXiv:1106.5857 [hep-ph]}

\bibitem{KRS07b}  K. Kowalski, J. Rembielinski, and K.A. Smolinski,
Phys.Rev. D \textbf{76}, 127701 (2007). arXiv:0712.2728

\bibitem{Kapusta89}  J. I. Kapusta and C. Gale, \textit{Finite-temperature
field theory, Principles and Applications}, 2nd ed. (Cambridge Univ. Press,
Cambridge, 2006), p. 8.

\bibitem{LL5}  L. D. Landau and E. M. Lifshitz, \textit{Statistical physics,
Part I }, 3rd ed. (Pergamon Press, Oxford, 1980), p. 47.

\bibitem{Ziman}  J. M. Ziman, \textit{Principles of the Theory of Solids}
(Cambridge Univ. Press, Cambridge, 1972), p. 137.

\bibitem{TV2011f}  E. Trojan and G. V. Vlasov, Thermodynamics of exotic
matter with constant w=P/E. {\tt arXiv:1108.0824 [cond-mat.stat-mech]}

\bibitem{Ziman2}  J. M. Ziman, \textit{Principles of the Theory of Solids}
(Cambridge Univ. Press, Cambridge, 1972), p. 163.
\end{thebibliography}
\end{document}